\documentclass[
,longbibliography%
,reprint%
]{revtex4-1}
\usepackage{graphicx}
\usepackage{txfonts}
\usepackage[english]{babel}

\usepackage{fancyhdr}
\newlength{\figsize}
\begin{document} 
\title
{Giant enhancement and sign inversion of optical Kerr nonlinearity in random high index nanocomposites near Mie resonances}
\author{Andrey V. Panov}
\email{Electronic mail: andrej.panov@gmail.com} 
\affiliation{ 
Institute of Automation and Control Processes,
Far East Branch of Russian Academy of Sciences,
5, Radio st., Vladivostok, 690041, Russia}



\begin{abstract}

High index dielectric nanoantennas excited at Mie-type resonances have exhibited enormous enhancement of optical nonlinearity.
Such nanostructures have been actively studied by researchers in recent years.
The present work provides the first 
numerical analysis study 
of the optical Kerr effect of nanocomposites consisting of high
refractive index (GaP) spheres at the wavelength of 532~nm.
This is done by means of 
3D finite-difference time-domain 
simulations.
The effective nonlinear refractive index of $0.8$~$\mu$m thick nanocomposites and metasurfaces is evaluated.
It is shown that the optical Kerr nonlinearity of the nanocomposites rises by orders in proximity to Mie resonances and may exceed the second-order refractive index of the bulk material. 
It is revealed that the sign of the effective optical Kerr coefficient is inverted near the Mie resonances.
This effect may be of interest in developing nonlinear optical metadevices.


\end{abstract}

\maketitle
\doi{10.1002/andp.201900574}

\section{Introduction}

During the past years, dielectric metasurface technology has been rapidly developed. 
The dielectric metasurfaces comprising interfaces patterned by high index particles of subwavelength size exhibit exceptional abilities for controlling light \cite{Genevet17}.
Nowadays, the shape and size of the high index particles can be precisely governed. 
For example, in Ref.~\cite{Verre18}, it was demonstrated the fabrication of the metasurfaces as monolayers of silicon nanoparticles of various shapes, sizes and compositions. 

In recent years, nonlinear optical properties of nanostructures containing dielectric high index Mie-type resonant particles have attracted much interest due to their potential applications for designing  light sources, optical modulators and novel ultrafast metadevices \cite{Smirnova16}. 
Near the resonances, the nanocomposites exhibit inherently large nonlinear response because of optical field localization in the nanoparticles. 
These particles with high refractive index show enhanced optical nonlinearity owing to the field concentration at the optical resonances \cite{Tribelsky16}. 
Mostly, researchers study elevated third harmonic generation in such nanocomposites. 
The efficiency of the harmonic generation with near resonant particles is enhanced by two orders of magnitude with respect to the bulk material \cite{Shcherbakov14}. 
Yang et al. \cite{Yang15} measured even higher increase in the third harmonic generation by a Fano-resonant silicon metasurface with the $1.5\times10^5$ factor in relation to an unpatterned Si film.
Also, the silicon nanoparticles exhibited Raman scattering elevated by two or three orders of magnitude at Mie-type resonances \cite{Cao06,Dmitriev16}.
Another nonlinear optical phenomenon, the optical Kerr effect, is frequently utilized to design all-optical switching compact devices.
However, the effective Kerr nonlinearity of the systems of dielectric Mie-type resonant particles has been left uninvestigated. 
This work provides the study intended to fill the gap.

Recently, there was proposed a method of restoration of the effective Kerr nonlinearity of nanocomposite media on the base of 
3D finite-difference time-domain (FDTD) simulations of light propagation \cite{Panov18}. 
This technique exploits the phase change induced by the studied sample to the transmitted Gaussian beam. 
This phase change is computed for different intensities of the Gaussian beam enabling one to estimate the real parts of the nonlinear refractive index of the sample.
The phase change is accounted at far distance from the sample on the beam axis so that the effect of the scattered in all  directions or multiple reflected irradiation on the phase at the axis is minimal.

In this work, the procedure presented in Ref.~\cite{Panov18} is applied to evaluation of the effective nonlinear refractive index of the random nanocomposite containing identical spherical high index inclusions with sizes close to the lowest Mie resonances.
The term ``effective" for inclusions with sizes close to the Mie resonances should not be understood in the sense of the effective medium. 
Less often it is designated as ``equivalent".
The disordered arrangement of nanoparticles employed here is expected to be less affected by mutual interplay as compared to lattices.
The effective second-order index of refraction is estimated at the light wavelength of 532~nm for the thick nanocomposites and the metasurfaces containing one layer of the spheres.

\section{FDTD simulation details}

\begin{figure*}
{\centering\includegraphics[width=14cm]{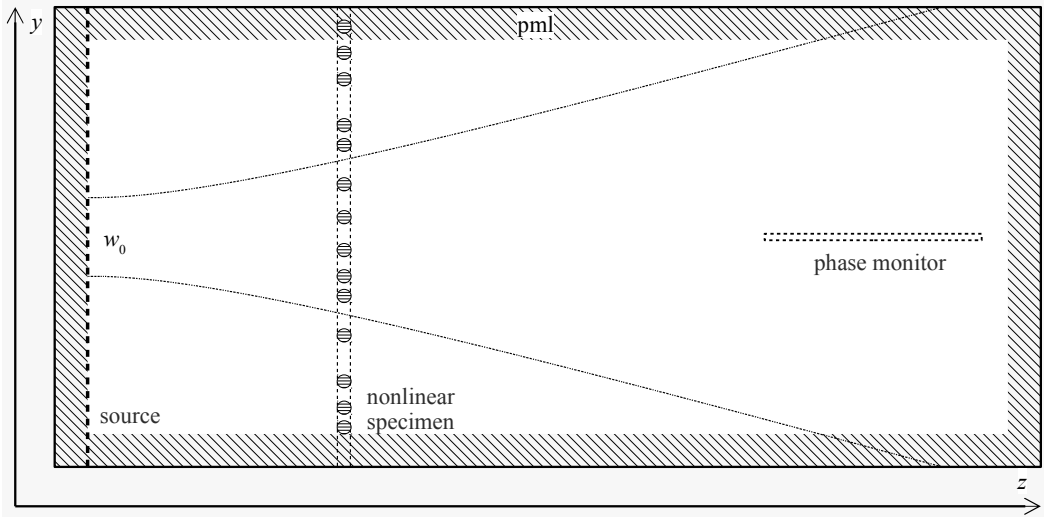}\par} 
\caption{\label{n2randfig} Schematic diagram of the 
3D FDTD simulation of the Gaussian beam propagation from the plane light source through a nonlinear nanocomposite. 
The instant electric field data gathered in points of the phase monitor allows one to calculate the phase shift introduced by the specimen. 
The phase shift carries the information about the nonlinear refractive index of the sample. 
There are perfect matched layers (PML) at the boundary of computational domain.}
\end{figure*}

The studied samples represent the random arrangement of the disjoint spheres of the same radius $r$ in space or on plane. 
There is a minimum distance between spheres.
The medium surrounding the spheres is vacuum. 
The procedure of random arrangement production prevents the overlap or touching of the neighboring spheres.
As depicted in Fig.~\ref{n2randfig}, the modeled Gaussian beam falls perpendicularly on the specimen with intensity-dependent index of refraction
\[
 n=n_0+n_2 I,
\]
where $n_0$ is the linear refractive index, $n_2$ is the second-order nonlinear refractive index, and $I$ is the intensity of the wave. 
Then, the phase change on the axis of the transmitted beam is calculated in several points in the phase monitor positioned far enough from the sample (see Fig.~\ref{n2randfig}). 
The phase change is computed by applying the discrete Fourier transform to the accumulated during simulations instant electric field component in the points of the phase monitor.
These simulations are conducted for different values of beam intensity $I$.
The linear fit of the phase change against the beam intensity provides the real part of the second-order index of refraction $n_2$. 
The nonlinear refractive index computed in the several points permits to estimate its average and standard deviation.

The gallium phosphide (GaP) has a high linear refractive index $n_{0\,\mathrm{in}}=3.49$ at wavelength $\lambda=532$~nm \cite{Aspnes83} and moderately low extinction coefficient (0.0026) which was neglected in this study. 
The  Massachusetts Institute of Technology (MIT) Electromagnetic Equation Propagation (MEEP) FDTD solver \cite{OskooiRo10} used in this work cannot simulate both nonlinear and lossy medium.
Thus, GaP was selected as a material of the inclusions for modeling since frequently used in experiments silicon has much larger absorption in the visible range. 
The third-order optical susceptibility of the bulk gallium phosphide was measured in the visible range as $\chi^{(3)}_{\mathrm{in}}\approx 2 \times 10^{-10}$~esu \cite{Kuhl85} that yields $n_{2\,\mathrm{in}}\approx 6.5\times10^{-17}$~m$^2$/W. 
The values of $n_2$ measured for the infrared wavelengths are one order of magnitude lower than the first estimate \cite{Martin18,Wilson18}. 
The first value was applied to the computations. 
In any case, the values of effective $n_{2\,\mathrm{eff}}$ calculated in the present work  can be scaled  for adoption to other magnitude of the second-order index of refraction of bulk GaP.

As the size of particles approaches the Mie resonances, the shape of the Gaussian beam transmitted through the specimen is considerably distorted so the size of the FDTD computational domain should be enlarged in comparison to Ref.~\cite{Panov18} in order to obtain the stable results. 
The size of the computational domain for simulations was $4\times 4\times 30$~$\mu$m with the space resolution of 5~nm. 
In the simulations, the distance between the Gaussian beam source and the studied sample was 0.9~$\mu$m, the beam radius $w_0$ at the beam waist was $1.1$~$\mu$m.

\section{Results and discussion}

\begin{figure*}
{\centering
\begin{minipage}[b]{\figsize}
\includegraphics[width=\figsize]{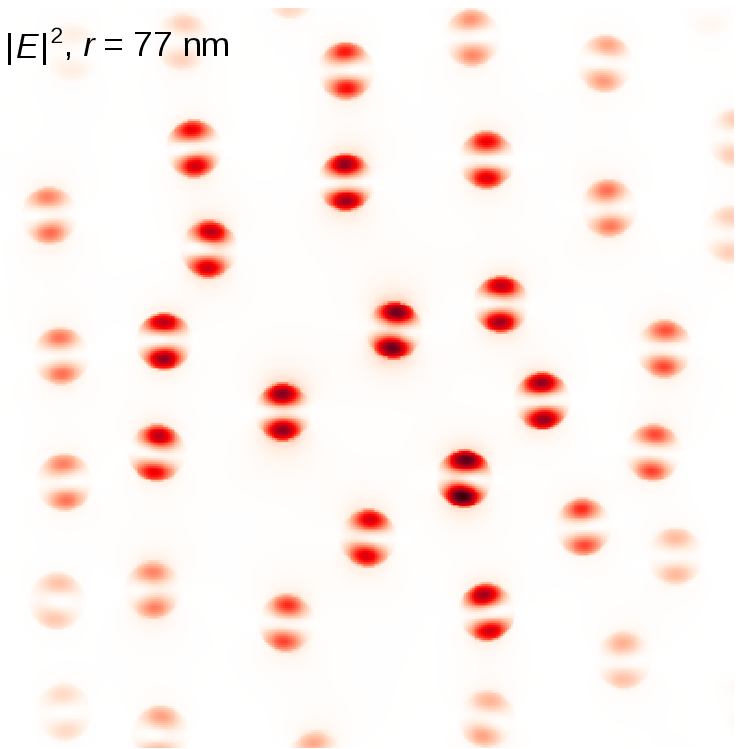}
\end{minipage}\hfill
\begin{minipage}[b]{\figsize}
\includegraphics[width=\figsize]{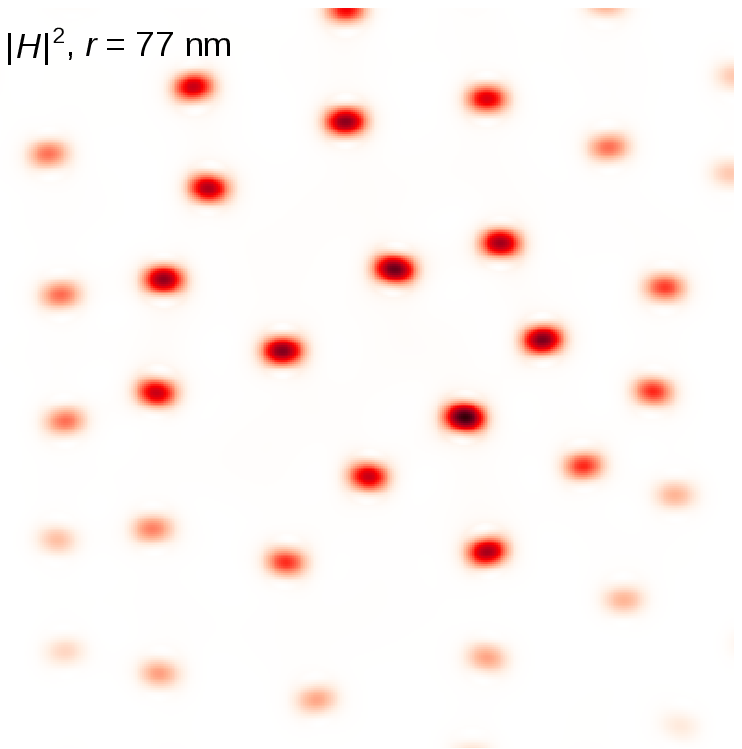}
\end{minipage}
\linebreak
\begin{minipage}[b]{\figsize}
\includegraphics[width=\figsize]{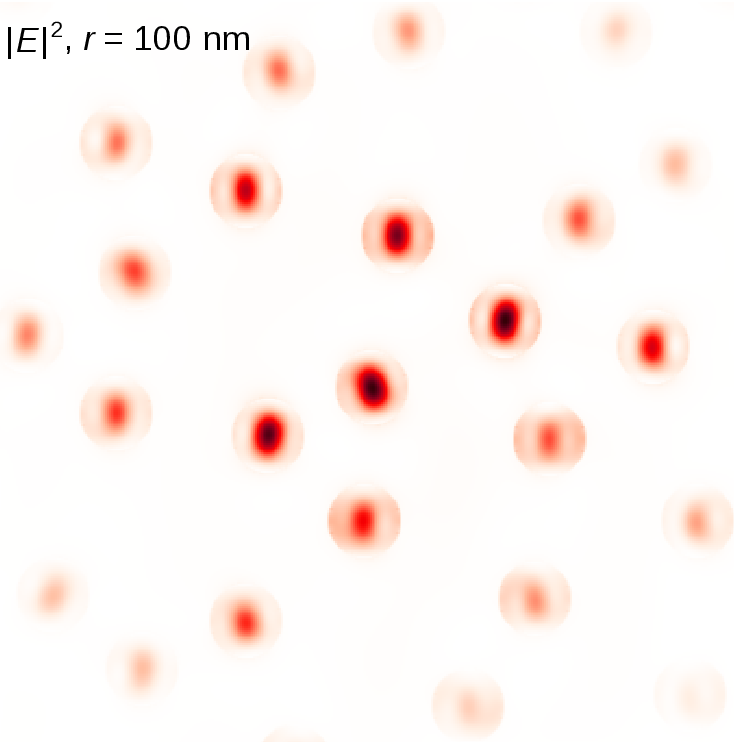}
\end{minipage}\hfill
\begin{minipage}[b]{\figsize}
\includegraphics[width=\figsize]{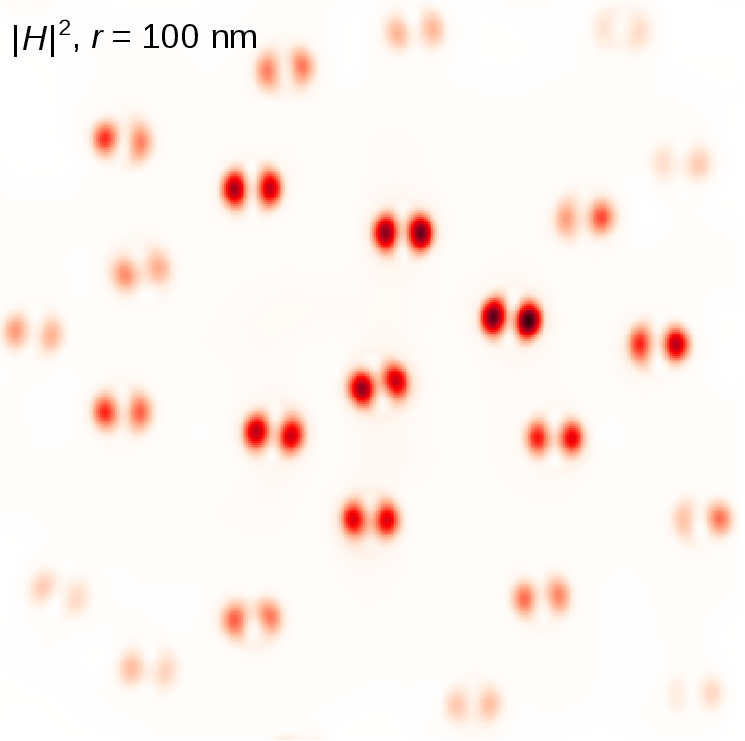}
\end{minipage}
\par} 
\caption{\label{ener_dist_mag_el_res} (Color online)
Central parts of time-average distributions of electric $|E|^2$ and magnetic $|H|^2$ energy densities in the disordered monolayers of GaP spheres at magnetic ($r=77$~nm) and electric ($r=100$~nm) dipole Mie resonances.
The distributions are calculated within the plane intersecting centers of the spheres.
The incident Gaussian beam is polarized along the vertical direction.}
\end{figure*}

At the Mie resonances, the small dielectric spheres with the high refractive index display modes with the self-main\-tai\-ning oscillations of the electric and magnetic fields. 
The first resonance in the dielectric particles with relative permittivity $\varepsilon>0$ is a magnetic dipole resonance.
The radii $r$ of the magnetic and electric resonances in spheres are given by \cite{Debye1909,Hulst81}
\[
\frac{2\pi r n_\mathrm{in}}{\lambda}=c_{j-1},\quad \frac{2\pi r n_\mathrm{in}}{\lambda}=c_j \left( 1 - \frac{1}{n_\mathrm{in}^2 j}\right) ,
\]
where $n_\mathrm{in}$ is the refractive index of inclusions, $j=1,2,\ldots$, $c_0=\pi$, $c_1=4.493$, $c_2=5.764$.
According to Debye formulas, 
the lowest magnetic dipole and quadrupole Mie resonances for the standalone GaP spheres at $\lambda=532$~nm are anticipated for the radii of $r=76$ and 109~nm, the first electric  resonances are expected at $r=100$~nm and 134~nm. 
These values may change significantly in a periodic lattice of the particles or at large concentrations of inclusions (see, e.g. \cite{Silveirinha11}). 
The distributions of electric and magnetic energy densities in the disordered monolayers of GaP spheres at the magnetic ($r=77$~nm) and electric ($r=100$~nm) dipole Mie resonances are presented in Fig.~\ref{ener_dist_mag_el_res}.
The energy distributions can be clearly identified as the magnetic and electric dipole Mie resonances.

\begin{figure}
{\centering
\includegraphics[width=\figsize]{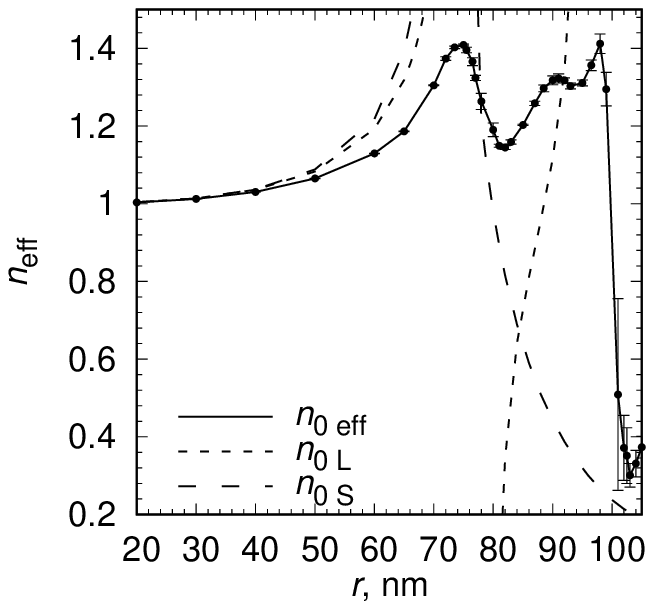}
\par} 
\caption{\label{n0_r_GaP_725}The effective linear refractive index of the $0.8$~$\mu$m thick random nanocomposite as a function of the sphere radius~$r$ for a fixed number of the particles (725). The effective refractive index $n_{0\,\mathrm{eff}}$ is calculated using FDTD modeling, $n_{0\,\mathrm{L}}$ and $n_{0\,\mathrm{S}}$ are results of the effective medium theories \cite{Lewin47,Slovick17}.}
\end{figure}

Fig.~\ref{n0_r_GaP_725} depicts the effective linear refractive index $n_{0\,\mathrm{eff}}$ of the $0.8$~$\mu$m thick disordered nanocomposite calculated using the FDTD simulations. 
The number of particles in the sample with the size of $4\times 4\times 0.8$~$\mu$m was fixed to 725, the radius of the spheres was varied. 
For comparison, the real parts of the effective index of refraction resulting from the effective medium formulas \cite{Lewin47,Slovick17} are shown. 
It is worth to be mentioned that Lewin's theory \cite{Lewin47} describes the cubic lattice of the spheres taking into consideration the first electric and magnetic Mie resonances. 
The similar effective medium approximation for the random distribution of dielectric spheres is developed in Ref.~\cite{Slovick17}. 
The dips of the curve  $n_{0\,\mathrm{eff}}(r)$ correspond to the Mie resonances.
The Mie resonances results in discontinuities of the effective permeability or permittivity depending on the type of the resonance \cite{Lewin47,Silveirinha11}.
The permeability (permittivity) tends to positive infinity below the resonance, then it increases from negative infinity above the resonance.
The radii of the Mie resonances are close to the values predicted by Debye formulas, just the first electric resonance is calculated slightly below 100~nm.
As can be seen from Fig.~\ref{n0_r_GaP_725}, the dependence $n_{0\,\mathrm{eff}}(r)$ obtained using FDTD simulations correlates with Lewin's theory while the first electric resonance is shifted by the model of Ref.~\cite{Slovick17}.
The discordance between these calculations and theory~\cite{Slovick17} may arise from an assumption of zero scattering from the nanocomposite, whereas it is doubtful for the resonant disordered metamaterial.
Theory~\cite{Slovick17} was tested in the limiting case of small inclusions where it reduces to the classic Bruggeman formula \cite{Bruggeman35} and for the diluted composite of the spheres with sizes far from the resonances transiting to Lewin's expressions.
Thus, the present study showed that, at least, the first magnetic and electric resonances are not changed appreciably for moderate volume fractions of the randomly positioned spheres.

\begin{table}[tb]
{
\renewcommand{\arraystretch}{1}
\centering 
\begin{tabular}{p{3ex}lll}
\hline
$r$, nm	&$f$, \% &$n_{0\,\mathrm{eff}}$&$n_{2\mathrm{eff}}$,~m$^2$/W	\\
\hline

20 &	0.2804 &	1.00334 &	$(5\pm2)\times 10^{-21}$ \\
20 &	4.591 &	1.0774 &	$(1.19\pm0.09)\times 10^{-19}$ \\
30 &	0.9063 &	1.01196 &	$(2.6\pm0.7)\times 10^{-20}$ \\
40 &	2.060 &	1.02929 &	$(1.1\pm0.2)\times 10^{-19}$ \\
50 &	3.860 &	1.0633 &	$(5.2\pm0.5)\times 10^{-19}$ \\
60 &	6.408 &	1.1256 &	$(4.8\pm0.3)\times 10^{-18}$ \\
65 &	7.988 &	1.1809 &	$(2.7\pm0.2)\times 10^{-17}$ \\
72 &	10.56 &	1.363 &	$(5.1\pm0.9)\times 10^{-16}$ \\
76 &	7.338 &	1.23 &	$(-4.8\pm0.7)\times 10^{-15}$ \\
78 &	7.870 &	1.050 &	$(-3.5\pm0.1)\times 10^{-16}$ \\
80 &	14.05 &	1.17 &	$(-6.8\pm1.6)\times 10^{-16}$ \\
80 &	14.05 &	1.18 &	$(-8.2\pm0.5)\times 10^{-16}$ \\
85 &	16.54 &	1.198 &	$(1.1\pm0.4)\times 10^{-16}$ \\
90 &	19.27 &	1.31 &	$(1.2\pm0.4)\times 10^{-16}$ \\
105 &	28.97 &	0.39 &	$(8.5\pm2.9)\times 10^{-16}$ \\
\hline
72 &	10.50 &	1.377 &	$(4.8\pm1.2)\times 10^{-16}$ \\
\hline

\end{tabular}

\par}
\caption{\label{n2_r_GaP_0.8mum}
The effective linear $n_{0\,\mathrm{eff}}$ and second-order $n_{2\,\mathrm{eff}}$ refractive indexes of the thick specimens reconstructed with the FDTD modeling 
for different radii of spheres $r$. 
Here $f$ is the volume fraction (concentration) of inclusions, $n_{2\,\mathrm{eff}}$  is given with the standard deviation. 
The last row was calculated for 1.6~$\mu$m thick sample, otherwise the thickness of the specimens was 0.8~$\mu$m.}
\end{table}

\begin{figure}
{\centering
\includegraphics[width=\figsize]{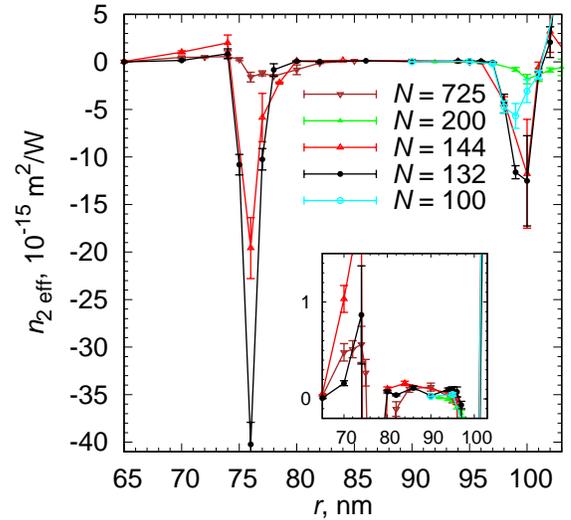}
\par} 
\caption{\label{n2_r_GaP_randsurf} (Color online) 
The effective second-order refractive index $n_{2\,\mathrm{eff}}$ of the disordered metasurfaces ($N=100, 132, 144$) and $0.8$~$\mu$m thick nanocomposites ($N=200, 725$) as a function of the sphere radius~$r$ for number $N$ of the particles in the specimens in the vicinity of the magnetic ($r\approx 76$~nm) and electric ($r\approx 100$~nm) dipole resonances. The inset shows the same plot in a magnified scale to emphasize the curves for lower values of $n_{2\,\mathrm{eff}}$.}
\end{figure}

Table~\ref{n2_r_GaP_0.8mum} and Fig.~\ref{n2_r_GaP_randsurf} provide the calculated values of the effective intensity-dependent refractive indexes of 0.8~$\mu$m thick samples with different radii of the spheres. 
The estimates of the effective optical Kerr nonlinearity of disordered metasurfaces are illustrated in Tab.~\ref{n2_r_GaP_surf} and Fig.~\ref{n2_r_GaP_randsurf}. 
The metasurfaces were monolayers of the particles of the same radius. The thickness of the metasurface was equal to the diameter of the sphere.
For some sizes of inclusions, several realizations of random arrangement was utilized for the FDTD simulations, these cases are seen in the tables as rows with the same particle radii. 
To validate, the 1.6~$\mu$m thick sample with spheres having the radius of 72~nm was modeled. 
The results are comparable with the 0.8~$\mu$m thick sample. 
The wide scatter of the retrieved magnitudes of the nonlinear refractive index of the metasurfaces with different arrangements of the 80~nm radius spheres may be associated with large variations of the particle density. 
These samples were modeled as 159 nanoparticles lying in the $4\times 4$~$\mu$m area.
The interparticle interactions are amplified in the vicinity of the resonance.

\begin{table}
{
\renewcommand{\arraystretch}{1}
\centering 
\begin{tabular}{p{3ex}lll}
\hline
$r$, nm	&$f$, \% &$n_{0\,\mathrm{eff}}$&$n_{2\mathrm{eff}}$,~m$^2$/W	\\
\hline
20 &	0.2825 &	1.00387 &	$(6.6\pm2.8)\times 10^{-21}$ \\
40 &	2.296 &	1.03542 &	$(1.6\pm0.3)\times 10^{-19}$ \\
50 &	3.896 &	1.0647 &	$(6\pm2)\times 10^{-19}$ \\
60 &	6.393 &	1.1400 &	$(8\pm2)\times 10^{-18}$ \\
65 &	7.938 &	1.218 &	$(5\pm1)\times 10^{-17}$ \\
70 &	9.791 &	1.371 &	$(1.0\pm0.1)\times 10^{-15}$ \\
70$^*$&	9.867 &	1.380 &	$(-1.2\pm0.1)\times 10^{-15}$ \\
70 &	9.791 &	1.372 &	$(9\pm1)\times 10^{-16}$ \\
72 &	10.58 &	1.541 &	$(4.3\pm0.8)\times 10^{-15}$ \\
72 &	10.55 &	1.515 &	$(3.8\pm0.5)\times 10^{-15}$ \\
80 &	14.01 &	0.926 &	$(-2.4\pm0.7)\times 10^{-16}$ \\
80 &	14.16 &	0.963 &	$(-5.5\pm0.9)\times 10^{-16}$ \\
80$^*$&	14.16 &	0.963 &	$(5.5\pm1)\times 10^{-16}$ \\
80 &	13.94 &	0.946 &	$(-4\pm1)\times 10^{-16}$ \\
85 &	16.66 &	1.023 &	$(9.8\pm1.4)\times 10^{-17}$ \\
90 &	19.27 &	1.115 &	$(1.5\pm0.1)\times 10^{-16}$ \\
105 &	28.86 &	0.352 &	$(9\pm5)\times 10^{-16}$ \\
\hline

\end{tabular}

\par}
\caption{\label{n2_r_GaP_surf}
The linear $n_{0\,\mathrm{eff}}$ and second-order $n_{2\,\mathrm{eff}}$ refractive indexes of disordered metasurfaces reconstructed with the FDTD modeling 
for different radii of spheres $r$. 
Here $f$ is the volume fraction of inclusions. 
Asterisk denotes artificial material of the inclusions with $n_{0\,\mathrm{in}}=3.49$ and $n_{2\,\mathrm{in}}=- 6.5\times10^{-17}$~m$^2$/W.}
\end{table}

\begin{figure}
{\centering
\includegraphics[width=\figsize]{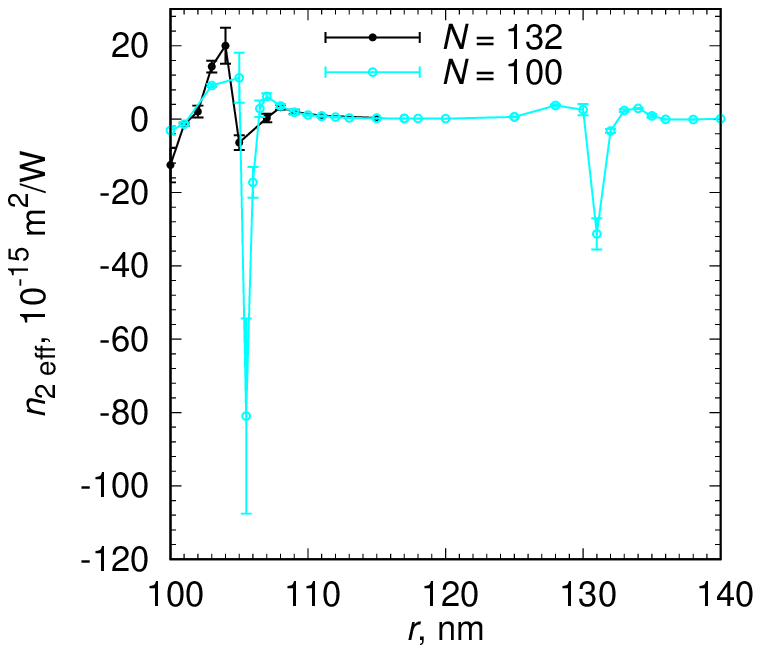}
\par} 
\caption{\label{n2_r_GaP_randsurf_quad} (Color online) 
The effective second-order refractive index $n_{2\,\mathrm{eff}}$ of the disordered metasurfaces ($N=100, 132$) as a function of the sphere radius~$r$ for number $N$ of the particles in the specimens in the vicinity of the magnetic ($r=105{-}108$~nm) and electric ($r\approx 132$~nm) quadrupole resonances.}
\end{figure}

It is particularly interesting that in the neighborhood of the optical resonances $n_{2\mathrm{eff}}$ is several orders of magnitude larger than would be expected from the non-resonant nonlinear effective medium theory and even exceeds the second-order refractive index of bulk GaP ($6.5\times10^{-17}$~m$^2$/W). 
This phenomenon arises from the giant field concentration inside the high index spheres close to the resonances \cite{Tribelsky16}. 
The optical Kerr nonlinearity of the nanocomposite is larger than that of the bulk material by one or two orders of magnitude near the Mie resonances. 
The similar behavior of the third harmonic generation from silicon nanodisks in proximity to the magnetic dipole resonance was observed in Ref.~\cite{Shcherbakov14}. 
It should be noted that the second-order refractive index of the metasurface in the range of the sphere sizes just below the magnetic dipole resonance is an order of magnitude higher than that of the thick samples with the same volume fraction of the nanoparticles.

\begin{figure}[tbh]
{\centering\includegraphics[width=\figsize]{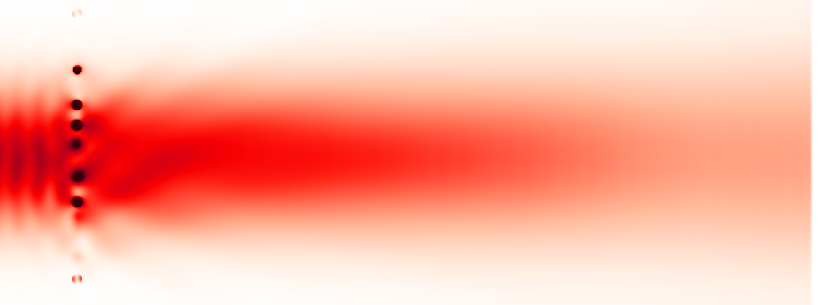}\par}
{\centering\includegraphics[width=\figsize]{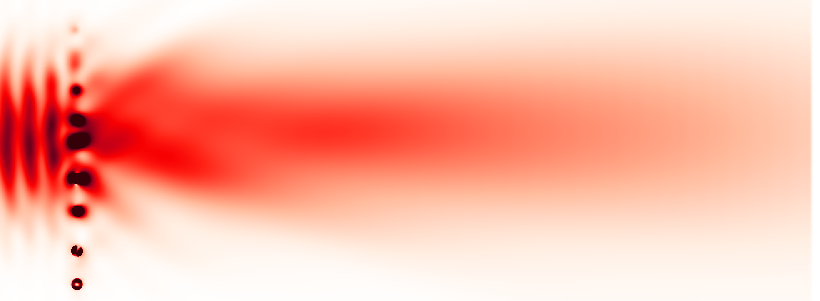}\par}
{\centering\includegraphics[width=\figsize]{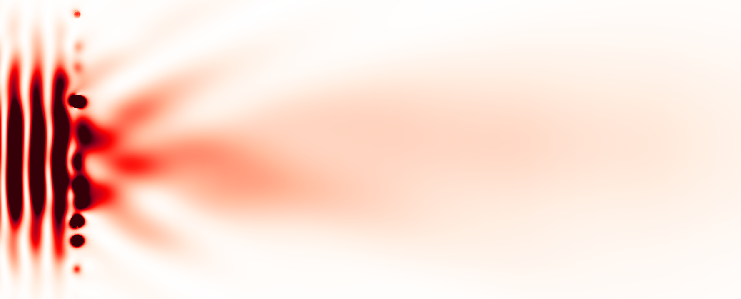}\par}
\caption{\label{elenertrans} (Color online) Distributions of the time-average electric energy density within the cross section of the initial part of the computational domain along the Gaussian beam axis for different radii of the spheres: 65~nm, 70~nm, 80~nm (from top to bottom). The beam is incident on a monolayer of the GaP spheres.}
\end{figure} 

Of special importance is the fact that the sign of the effective second-order refractive index is inverted in some range of the sphere sizes close to the magnetic dipole resonance ($r=76{-}82$~nm) or the first electric resonance ($r=97{-}101$~nm) for sparsely packed samples. 
This phenomenon  emerges both for the thick specimens and the metasurfaces. 
Figs.~\ref{n2_r_GaP_randsurf}, \ref{n2_r_GaP_randsurf_quad} illustrate the dependencies of $n_{2\,\mathrm{eff}}$ for monolayers or $0.8$~$\mu$m thick arrangements of the spheres with fixed values of the particle number on their radii.
As obvious from Figs.~\ref{n2_r_GaP_randsurf}, \ref{n2_r_GaP_randsurf_quad} the effective second-order refractive index is enhanced just below the resonances, then $n_{2\,\mathrm{eff}}$ drops changing its sign and rises again.
The sizes of the electric dipole and magnetic quadrupole resonances are close and the metasurfaces show a peak of  $n_{2\,\mathrm{eff}}$ between them.
In order to validate the inversion of the $n_{2\,\mathrm{eff}}$ sign, two simulations of the material with $n_0$ of GaP and the negative sign of the second-order refractive index were performed (marked with asterisk in Tab.~\ref{n2_r_GaP_surf}). 
In these cases, in the range of particle sizes near the magnetic dipole resonance $n_{2\,\mathrm{eff}}$ is positive. 
As a possible reason for this  this phenomenon, it should be considered negative values of effective magnetic permeability $\mu_\mathrm{eff}$ (or electric permittivity $\varepsilon_\mathrm{eff}$) which are observed for the inclusion sizes above the magnetic (or electric) resonances (see, e.g., \cite{Holloway03,Silveirinha11}). 
Under these circumstances, the specimen serves as a single negative metamaterial. 
When $\mu_\mathrm{eff}<0$ or $\varepsilon_\mathrm{eff}<0$ and losses inside the sample are low, the light partially transmits through the specimen by evanescent tunneling. 
For sphere sizes somewhat above the Mie resonances, $n_{0\,\mathrm{eff}}<1$ was found. 
It is a known fact that the effective refractive index at Mie-type resonance may be under unity \cite{OBrien02}. 
In support of the hypothesis of single negative metamaterial is the fact that the $n_{2\,\mathrm{eff}}$ of the disordered bidisperse metasurfaces with negative effective refractive index with 77 and 101~nm radii of spheres is positive \cite{Panov20a}.

\begin{figure}[tbh]
{\centering
\includegraphics[width=\figsize]{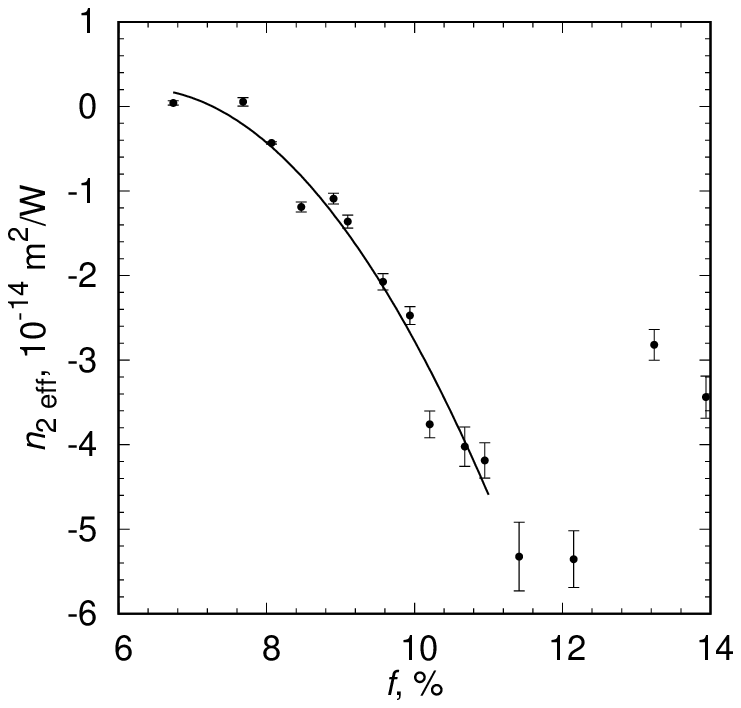}
\par} 
\caption{\label{n2_conc_magndip_GaP_surf} 
The peak effective second-order refractive index $n_{2\,\mathrm{eff}}$ of the disordered metasurfaces at the magnetic dipole resonance as a function of the volume fraction.}
\end{figure} 

Yet another possible explanation for the observed $n_{2\,\mathrm{eff}}$ sign inversion might be the effect of interparticle interactions on the nonlinear properties of the metasurface.
From simple geometric considerations, the number of two-particle correlations is proportional to $N^2$ and, as a consequence, to the square of volume fraction $f^2$.
The two-particle correlations will prevail for the dilute nanocomposites.
Conversely, the reasons lying within the particles should cause a linear change in the net nonlinear effect with the particle concentration.
On the other hand, when $f\to 1$, $n_{2\,\mathrm{eff}}$ should approach the bulk value.
In order to examine the assumption of the interparticle interaction effect, the dependence of $n_{2\,\mathrm{eff}}$  at the magnetic dipole resonance of the metasurface on the volume fraction was constructed (Fig.~\ref{n2_conc_magndip_GaP_surf}). 
The points in this plot correspond to maximum values of $|n_{2\,\mathrm{eff}}|$ observed at $r=76$~nm (lower concentrations) or $r=77$~nm (higher concentrations) since the radius of the resonance is shifted at large volume fractions of the nanoparticles.
The line in Fig.~\ref{n2_conc_magndip_GaP_surf} shows the fit by a quadratic polynomial for $f<11$~\%.
This fit seems to follow the calculated dots but due to their scatter in the plot the assumption of the interparticle interaction effect could not be certainly proved.
The phenomenon of the $n_{2\,\mathrm{eff}}$ sign inversion requires further investigations while it may be of considerable significance for designing all-optical switches.

It should be emphasized that the only a minor part of the light intensity transmits through the single negative material (see Fig.~\ref{elenertrans}). 
When the sphere radius is 80~nm the input flux is mostly reflected forming a standing wave before the specimen. 
Moreover, the substantial portion of the incident energy is scattered by the particles in different directions in the vicinity of the resonance.
Whereas for the sphere radii just below magnetic dipole resonance, the nonlinear nanocomposite is much more transparent to light simultaneously having large magnitudes of $n_{2\,\mathrm{eff}}$. 
According to formulas 35 and 36 from Ref.~\cite{Tzarouchis18}, the first Kerker condition (maximum forward scattering from spheres) for GaP at $\lambda=532$~nm should occur for $r=67$~nm and the second Kerker condition (maximum backward scattering) will be observed for $r=69$~nm.
Thus, this range of particle sizes is preferable for designing nonlinear nanocomposites.

\section{Summary}

In summary, the real part of the effective Kerr nonlinear refractive index of the random nanocomposites consisting of the high index spheres is estimated. 
The dependence of the effective second-order refractive index of disordered metasurfaces on the sphere size is computed in proximity to the lowest magnetic and electric resonances.
It is shown that the nanostructure second-order refractive index near the Mie resonances exceeds that of the bulk material by one or two orders of magnitude. 
The sign of the second-order refractive index is inverted for the nanocomposite with the sphere diameters in the close vicinity of the sizes of the Mie resonances.

\subsection*{Acknowledgments}
The results were obtained with the use of IACP FEB RAS Shared Resource Center ``Far Eastern Computing Resource'' equipment (https://www.cc.dvo.ru).

\subsection*{Conflict of Interest} The author declares no conflict of interest.

\subsection*{Keywords}
dielectric nanoparticles, electromagnetic field localization, magnetic dipole resonance, optical Kerr nonlinearity, random nanocomposites

\bibliography{nlphase}

\end{document}